\documentclass[12pt,a4paper,final]{iopart}

\usepackage{color}
\usepackage{iopams}  
\usepackage{graphicx}
\usepackage{graphicx}
\usepackage[utf8x]{inputenc}
\usepackage[british]{babel}
\usepackage[margin=2cm]{geometry}
\usepackage{lastpage}
\usepackage{color}
\usepackage{eurosym}
\usepackage{rotating}
\usepackage{graphics}
\usepackage{cite}

\begin{document}

\title{Optimization of the geometrical stability in square ring laser gyroscopes}
\author{R.~Santagata$^{1,2}$, A.~Beghi$^{3}$, J.~Belfi$^{2}$, N.~Beverini$^{2,4}$, D.~Cuccato$^{3,5}$, A.~Di Virgilio$^2$, A.~Ortolan$^{6}$, A.~Porzio$^{7,8}$, and S.~Solimeno$^{9}$}

\address{$^1$Department of Physics, University of Siena, Via Roma 56, Siena, Italy}
\address{$^2$INFN Section of Pisa, Largo Bruno Pontecorvo 3, Pisa, Italy}
\address{$^3$Department of Information Engineering, University of Padova, Via Gradenigo 6/B, Padova, Italy}
\address{$^4$Department of Physics, University of Pisa, Largo Bruno Pontecorvo 3, Pisa, Italy}
\address{$^5$INFN Sezione di Padova, Via Marzolo 8, 35131, Padova, Italy}
\address{$^6$INFN National Laboratories of Legnaro, Viale dell'Universit\`a 2, Legnaro, Padova, Italy}
\address{$^7$CNR-SPIN Section of Napoli, Complesso Universitario Monte Sant'Angelo, Via Cintia, Napoli, Italy}
\address{$^8$INFN Section of Napoli, Complesso Universitario Monte Sant'Angelo, Via Cintia, Napoli, Italy}
\address{$^9$Department of Physics, University of Napoli, Complesso Universitario Monte Sant'Angelo, Via Cintia, Napoli, Italy}
\ead{\mailto{rosa.santagata@pi.infn.it}}

\begin{abstract}
Ultra sensitive ring laser gyroscopes are regarded  as potential detectors of the general relativistic frame-dragging effect due 
to the rotation of the Earth: the project name is GINGER (Gyroscopes IN GEneral Relativity), 
a ground-based triaxial array of ring lasers aiming at measuring the Earth rotation rate with an accuracy of $10^{-14}$ rad/s. Such ambitious goal is now within reach as large area ring lasers are very close to the necessary sensitivity and stability.  However, demanding constraints on the geometrical stability of the laser optical path inside the ring cavity are required. Thus we have started a detailed study of the geometry of an optical cavity, in order to find a control strategy for its geometry which could meet the specifications of the GINGER project. As the cavity perimeter has a stationary point for the square configuration, we identify a set of transformations on the mirror positions which allows us to adjust the laser beam steering to the shape of a square.  We show that the geometrical stability of a square cavity strongly increases by implementing a suitable system to measure the mirror distances, and that the geometry stabilization can be achieved by measuring the absolute lengths of the two diagonals and the perimeter of the ring. 

\end{abstract}

\section{Introduction}
A ring laser (RL) gyroscope is composed by a square or triangular
optical cavity inside which two opposite light beams propagate in a closed loop. In a reference frame rotating at velocity $\mathbf{\Omega}$ with respect to an inertial frame, the opposite directions are not
equivalent; the two laser emission frequencies are then splitted by the Sagnac
frequency $f_{S}$:
\begin{equation}
f_S=\frac{4 \mathbf{A}\cdot \mathbf{\Omega}}{\lambda p}=\mathbf{k_S} \cdot \mathbf{\Omega},
\end{equation}
where $\mathbf{k}_{S}=4\mathbf{A}/(\lambda p)$ is the oriented scale
factor, $\mathbf{A}$ is the vectorial area enclosed by the laser beams, $\lambda$ is the optical wavelength and $p$ the length of the round-trip path. 
The frequency difference can be measured by observing the beat note between the two counter-propagating beams. 

The fundamental limit to the angular velocity resolution of a RL is given by the photon shot-noise. 
The power spectral density of shot noise, converted in equivalent rotational noise and expressed
in unit of $\rm{(rad/s)/\sqrt{Hz}}$, reads \cite{StedmanShot}

\begin{equation}
P_{\Omega}^{1/2}=\frac{c}{\lambda || \mathbf{k}_{S} || Q}\sqrt{\frac{h\nu}{P_{out}T}}\;,\label{eq:ShotNoise} 
\end{equation}
where $Q$ is the quality factor of the optical cavity, 
$h$ the Planck constant, $P_{out}$ the optical power detected by photodiode,
$T$ the measuring time, $c$ the speed of light, and $\nu=c/\lambda$
the light frequency. 

Significant advancements in RL sensitivity have recently driven new
applications concerning Geophysics, Geodesy and test of General Relativity \cite{ullireview,JapanEarthquake,G-Chandler, PRD2011}.
A present challenge of this research is the observation of the relativistic Lense-Thirring effect, a perturbation 
of few parts in $10^{10}$ to the Earth rotation rate. Such a measurement requires a calibration of the RL with respect to the local spacetime, in order
to measure the Earth rotation with high accuracy. 
On the other hand, the time dependent fluctuation of scale factor, backscattering
and laser dynamics induced non-reciprocal effects have to be controlled in order to fully exploit the RL sensitivity up
to the shot noise level in equation~\ref{eq:ShotNoise}. In recent papers \cite{Cuccato,Beghi} we addressed the role of laser dynamics on the stability and accuracy 
of a ring laser
used as rotation sensor; in this work we focus on the role of cavity deformations 
on $\left\Vert \mathbf{k}_{S}\right\Vert $ and $p$.

The design of the cavity is crucial for the stability of the geometrical scale factor $\mathbf{k}_{S}$. The most stable and sensitive rotation sensor
at present, the Gross Ring G (Wettzell, Bavaria) \cite{GWettzell}, has an 
optical cavity with a side length of 4 m, and has been constructed on a rigid frame made of $\rm Zerodur^{\textregistered}$. In this case the geometrical stability is obtained by fixing the mirrors to a very rigid and heavy structure, and by minimizing thermal expansions and pressure variations. This experimental set-up has shown that ring lasers have the potentiality to measure the Lense-Thirring effect on an Earth based experiment, but the 'monolithic' design  cannot be extended to a triaxial array of ring laser with sides of  $6-10$ $\rm{m}$, as required by the GINGER project \cite{PRD2011}. An heterolithic design must be considered: the cavity frame can be made of more standard materials with thermal expansion coefficients at the level of several  \rm{ppm/K}, and the mirror positions
must be actively controlled in order to compensate the frame deformations by implementing a suitable system to measure the mirror distances 
based on ultra-stable optical frequency references. 

 A detailed study of the effects of geometrical distortions on the
optical path is the first necessary step toward the active control of mirror positions. 
In the literature, one can find many papers about the geometry of a ring cavity \cite{Cinesi1,Cinesi2,Cinesi3,Hurst-Geometry,Stedman,Bilger,Rodloff}.
In this paper we adress for the first time the problem to find a suitable approach to minimize the scale factor $\textbf{k}_{S}$ variations, and pose the basis for the control of an heterolithic ring laser cavity. Our study is based on  Fermat's principle to model the optical cavity geometry 
starting from the position of the center of curvature of the mirrors. 
We demonstrate that in a square cavity the control of the length of the two diagonals plays an important role. In fact, if the two RL diagonals are locked to the same absolute length, the optical cavity length has a stationary point corresponding to the regular square cavity, with two important implications for  the control of an heterolithic structure: i) one can implement a procedure to approach the regular geometry; ii) in this case, the residual deformations  contribute with quadratic terms depending on the ratio between the cavity side-length and the mirrors radius of curvature. 
\\
The paper is organized as follows. In section~2 we used the Fermat's
principle to calculate the light path inside a closed optical cavity formed
by four equal spherical mirrors. Section~3 is devoted to the decomposition of the 
cavity deformations in a suitable eigenvectors basis and to the estimation of their contributions 
to the cavity perimeter and scale factor. 
Section~4 reports the analysis of the perimeter and scale factor variations 
due to the residual deformations in a square cavity with fixed diagonal lengths; 
conclusions are drawn in section~5.

\section{The geometrical model of the optical cavity}
We consider a square ring optical cavity defined by four spherical mirrors. 
Let $r_{k}\in\mathbb{R}^{+}$
and $\mathbf{c}_{k}\in\mathbb{R}^{3}$ denote the radius and the center
of curvature of the $k_{th}$ mirror with $k=\{ 1,2,3,4\} $.
The four light spots can be usefully represented as the elements $X$
of $\mathbb{S}^{2 \times 4}$, i.e. $4$ points on the spheres representing
the mirrors; let $\mathbf{s}_{k}=r{}_{k}\mathbf{x}_{k}+\mathbf{c}_{k}$
be the position vector of the $k_{th}$ spot with respect to a reference
frame, and $\mathbf{\mathbf{x}_{\mathrm{k}}}\in\mathbb{S}^{2}$$=\{ \mathbf{x}\in\mathbb{R}^{3},\ \mathbf{x}^{T}\mathbf{x}=1\} $
the directional cosines of the spot position with respect to its mirror
center. We introduce the compact notation for the directional cosine matrix
$X=[\mathbf{x}_{1},\ldots,\mathbf{x}_{4}]\in\mathbb{S}^{2 \times 4}\subset\mathbb{R}^{3\times4}$,
the mirror configuration matrix $C=[\mathbf{c}_{1},\ldots,\mathbf{c}_{4}]\in\mathbb{R}^{3\times4}$,
and the curvature radii matrix $R=diag[r_{1},\dots,r_{4}]\in\mathbb{R}^{4\times4}$.
In this notation the spot position matrix can be written as $S=XR+C$.

Our aim is derive a functional dependence between the spot position matrix $S$ and the mirror configuration matrix $C$. Since each mirror displacement 
can be described as a variation of the position of its curvature center,
the four vectors $\mathbf{c}_{k}$ represent a set of dynamical
variables of the system. All the relevant quantities, with respect to the Sagnac signal and its noise, will be given in terms of these variables.
The 4 geometrical constraints $\Vert \mathbf{s}_{k}-\mathbf{c}_{k}\Vert ^{2}=r{}_{k}^{2}$,
$k=1,\dots,4$ ($\mathbf{x}_{k}$  are points of spherical surfaces),
reduce the dimensions of the set of $X$  from dim$(\mathbb{R}^{3\times4})=3\times4=12$
to dim$(\mathbb{S}^{2\times4})=2\times4=8$.

The Euclidean distances among the $6$ pairs $(k,j)$ of the four light spots are given by 
\begin{eqnarray}
\label{eq:length}
l_{kj}(X;C,R) & = & ||\mathbf{s}_{k}-\mathbf{s}_{j}|| \nonumber \\
 & = & ||r{}_{k}\mathbf{x}_{k}-r_{j}\mathbf{x}_{j}+\mathbf{c}_{k}-\mathbf{c}_{j}|| , \label{eq:distance}
\end{eqnarray}
where $||\cdot||$ is the Euclidean norm in $\mathbb{R}^{3}$. The optical path length $p$ 
is simply the sum of the lengths of the sides of the, in general non-planar, optical cavity
\begin{eqnarray}
p(X;C,R)& = & l_{12}(X;C,R)+l_{23}(X;C,R) \nonumber \\ 
& +& l_{34}(X;C,R)+l_{41}(X;C,R) \nonumber \ ,
\end{eqnarray} and the modulus of the enclosed area reads
\begin{eqnarray}
\Vert \mathbf{A}(X,C,R)\Vert
& =\frac{1}{2} & \Vert (\mathbf{s}_{1}-\mathbf{s}_{3})\wedge(\mathbf{s}_{2}-\mathbf{s}_{4})
\Vert =\label{eq:arearing} \nonumber \\
 & =\frac{1}{2} & l_{13}(X;C,R)\, l_{24}(X;C,R)\cos\vartheta\ ,
\end{eqnarray} where $\vartheta$ is the angle between the vectors $\mathbf{s}_{1}-\mathbf{s}_{3}$
and \textbf{$\mathbf{s}_{2}-\mathbf{s}_{4}$}. If the 4 mirror centers
lie in the same plane these vectors represent the diagonals of a polygon.
In this case $p(X;C,R)$,
$l_{13}(X;C,R)$ and $l_{24}(X;C,R)$ coincide with the perimeter
and diagonal lengths of the corresponding planar polygon.\\

\begin{figure}[h]
\centerline{\includegraphics[width=.7\columnwidth]{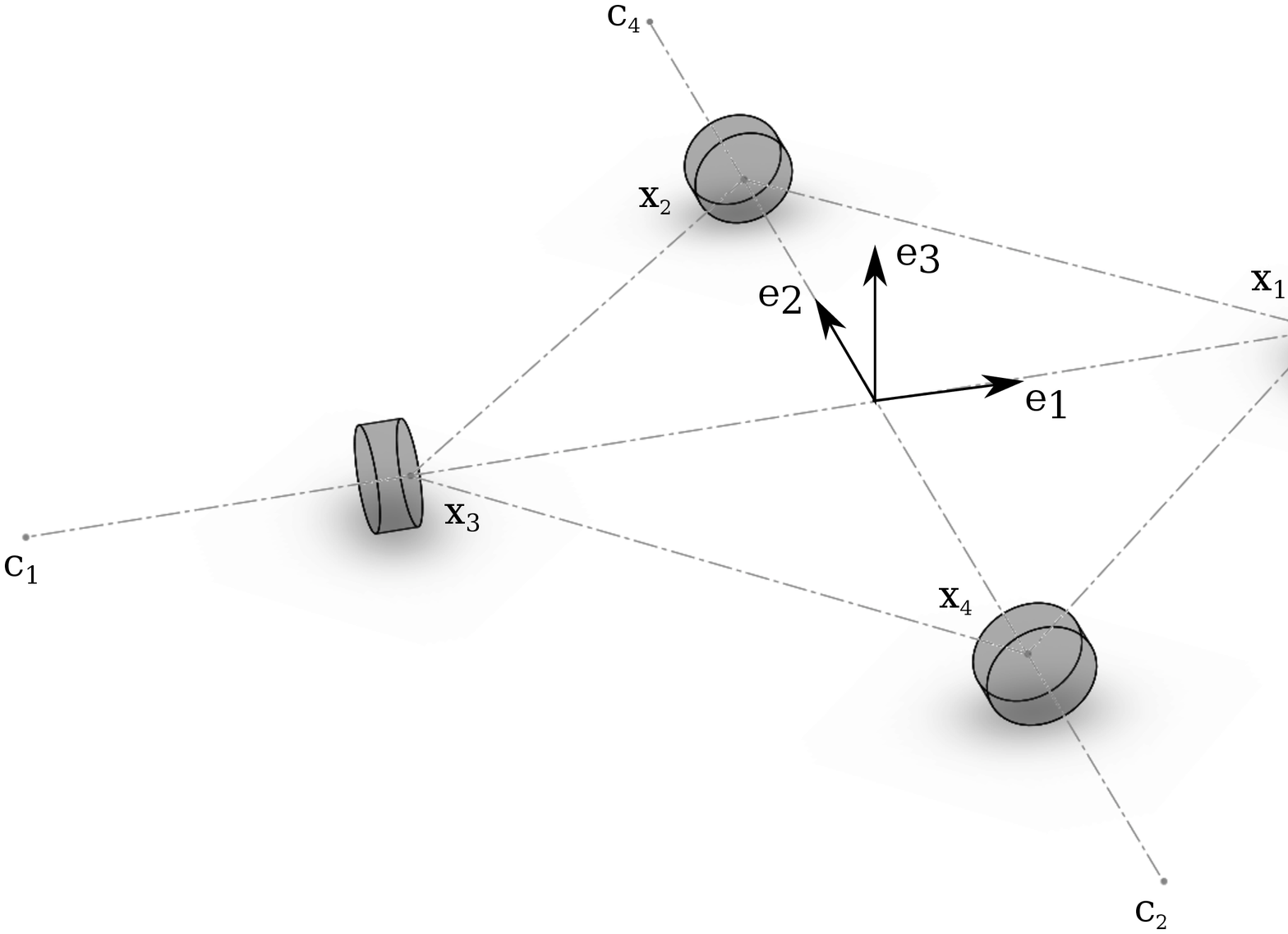}}
\caption{\label{fig:disegnocav} Schematic of the optical cavity. The position
of the $k_{th}$ mirror is determined by the coordinates of its center
of curvature $\mathbf{c}_{k}$. The light spots $\mathbf{x}_{k}$
are calculated by the Fermat's principle. The reference system has
been chosen with the origin in the center of the square, the axes
$\mathbf{e}_{1}$ and $\mathbf{e}_{2}$ along the two diagonals, and
$\mathbf{e}_{3}$ perpendicular to the square.}
\end{figure}

The effective directional cosine matrix $X(C,R)^{*}$ for a given
mirrors configuration $C$ and curvature radii $R$ is found by
using the Fermat's principle \cite{BornWolf} which states that light rays
follow extremal optical paths, i.e. 

\begin{equation}
\nabla p(X;C,R)=\mathbf{0}\ ,\label{eq:min}
\end{equation} where $\nabla(\cdot)$ is the gradient with respect to a coordinate set of $X\in \mathbb{S}^{2\times4}$. 
The vector components are referred 
to a clockwise orthonormal coordinate system
$\left(\mathbf{e}_{1},\mathbf{e}_{2},\mathbf{e}_{3}\right)$ with
the origin in the center of the square, the axes $\mathbf{e}_{1}$
and $\mathbf{e}_{2}$ along the two diagonals, and $\mathbf{e}_{3}$
perpendicular to the square (see figure~\ref{fig:disegnocav}).

In this coordinate system, the ideal mirrors configuration for the regular square cavity, 
with side length $L$ and all the mirror curvature radii equal to
$r$, is represented by the matrix
\begin{equation}
C_{0}= r\, \left (1-\frac{L}{\sqrt{2}r} \right )\left(\begin{array}{cccc}
1 & 0 & -1 & 0 \nonumber \\
0 & 1 & 0 & -1 \nonumber \\
0 & 0 & 0 & 0 \nonumber 
\end{array}\right)\ .
\end{equation}
Then the matrix
\begin{equation}
X_{0}=\left(\begin{array}{cccc}
-1 & 0 & 1 & 0 \nonumber \\
0 & -1 & 0 & 1 \nonumber \\
0 & 0 & 0 & 0 \nonumber
\end{array}\right)\ ,
\end{equation} is the solution of equations~\ref{eq:min} for the spot directional cosines. In general, equations~\ref{eq:min} are 
an irrational algebraic system
of equations, and no analytic solution is available. However, for small deformations, 
we can calculate the Taylor expansion of $p(X;C,R)$ in a neighborhood of $X_{0}$. \\

Let us consider the matrix $Z=X_{0}+Y$. The columns of $Y$ parametrize the tangent vectors to $X_{0}$, i.e.
each columns of $Y$, identified as a vector of $\mathbb{R}^{3}$,
is orthogonal to the corresponding column of $X_{0}$: 
\begin{equation}
Y=\left(\begin{array}{cccc}
0 & y_{3} & 0 & y_{7} \nonumber \\
y_{1} & 0 & y_{5} & 0 \nonumber \\
y_{2} & y_{4} & y_{6} & y_{8} \nonumber
\end{array}\right).
\end{equation}
We obtain the new point $X^{'}$ of $\mathbb{S}^{2\times4}$ using
the map $\mathcal{R}:\mathbb{S}^{2}\times\mathbb{R}^{3}\rightarrow\mathbb{S}^{2}$,
$\mathcal{R}(\mathbf{x},\mathbf{a})=(\mathbf{x}+\mathbf{a})/||\mathbf{x}+\mathbf{a}||$
column-wise on $Z$, so that each column of $X^{'}$ is the corresponding column of $Z$ normalized. The matrix $Z$ can be considered as a function
of the vector $\mathbf{y}=(y_{1},\dots,y_{8})^{T}$, and
therefore $X^{'}$ can be also regarded as a function of $\mathbf{y}$. A second order 
geometric model of the function $p(X^{'}(\mathbf{y});C,R)$
is then
\begin{equation}
p(X^{'}(\mathbf{y});C,R)=p(X_{0};C,R)+\mathbf{s}(C,R)^{T}\mathbf{y} \label{eq:model2}+\frac{1}{2}\mathbf{y}^{T}F(C,R)\mathbf{y}+o(||\mathbf{y}||^{2})\,, 
\end{equation}
where the vector $\mathbf{s}(C,R)$ and the matrix $F(C,R)$ are the gradient 
and Hessian of the function $p$ with respect to $\mathbf{y}$, respectively. 

Thus substituting equation~\ref{eq:model2} in equations~\ref{eq:min}
we find that the stationary point for this geometric model is given by
\begin{eqnarray}
\mathbf{y}^{*} & = & -F(C,R)^{-1}\mathbf{s}(C,R)\,.\label{eq:y_sol}
\end{eqnarray}
The corresponding extremal optical length is 
\begin{equation}
p^{*}(C,R)=p(X_{0};C,R)-\frac{\mathbf{s}(C,R)^{T}F^{-1}(C,R)\mathbf{s}(C,R)}{2}\ , \label{eq:p(C)} \nonumber
\end{equation}
 and the new spot positions are $S(C,R)^{*}=X^{'}\mathbf{(y}^{*})R+C$.
The matrix $F^{-1}(C,R)$ has a straightforward interpretation as
the ray matrix of the RL cavity, while the vector $\mathbf{y}$ can
be interpreted as the linear response of the optical cavity to a displacement
$\delta C$ of the centers of curvature from the ideal mirrors configuration.\newline

\section{The eigenvectors basis of the cavity deformations}
\begin{figure}[h!!]
\centerline{\includegraphics[width=5cm,height=5cm]{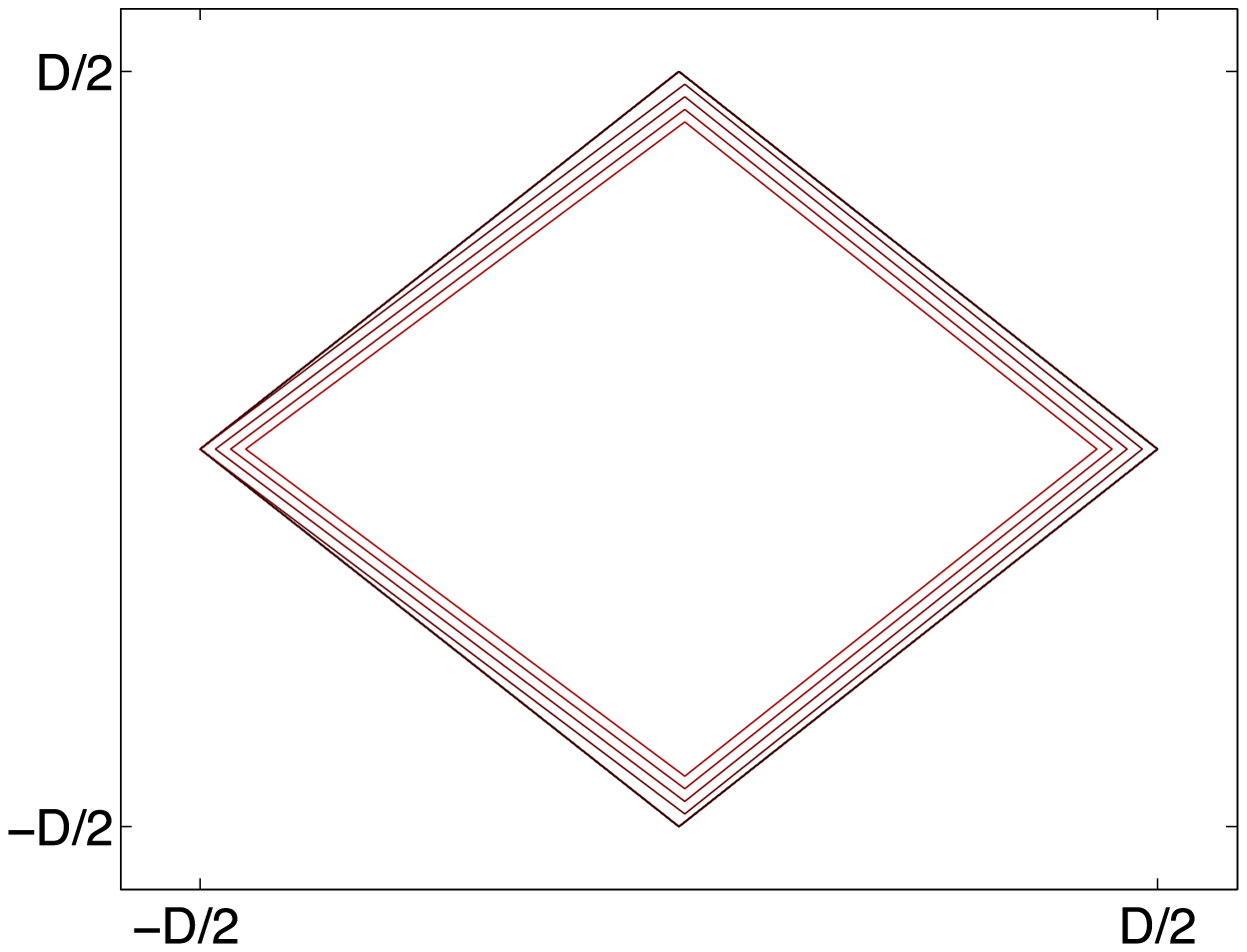},\includegraphics[width=5cm,height=5cm]{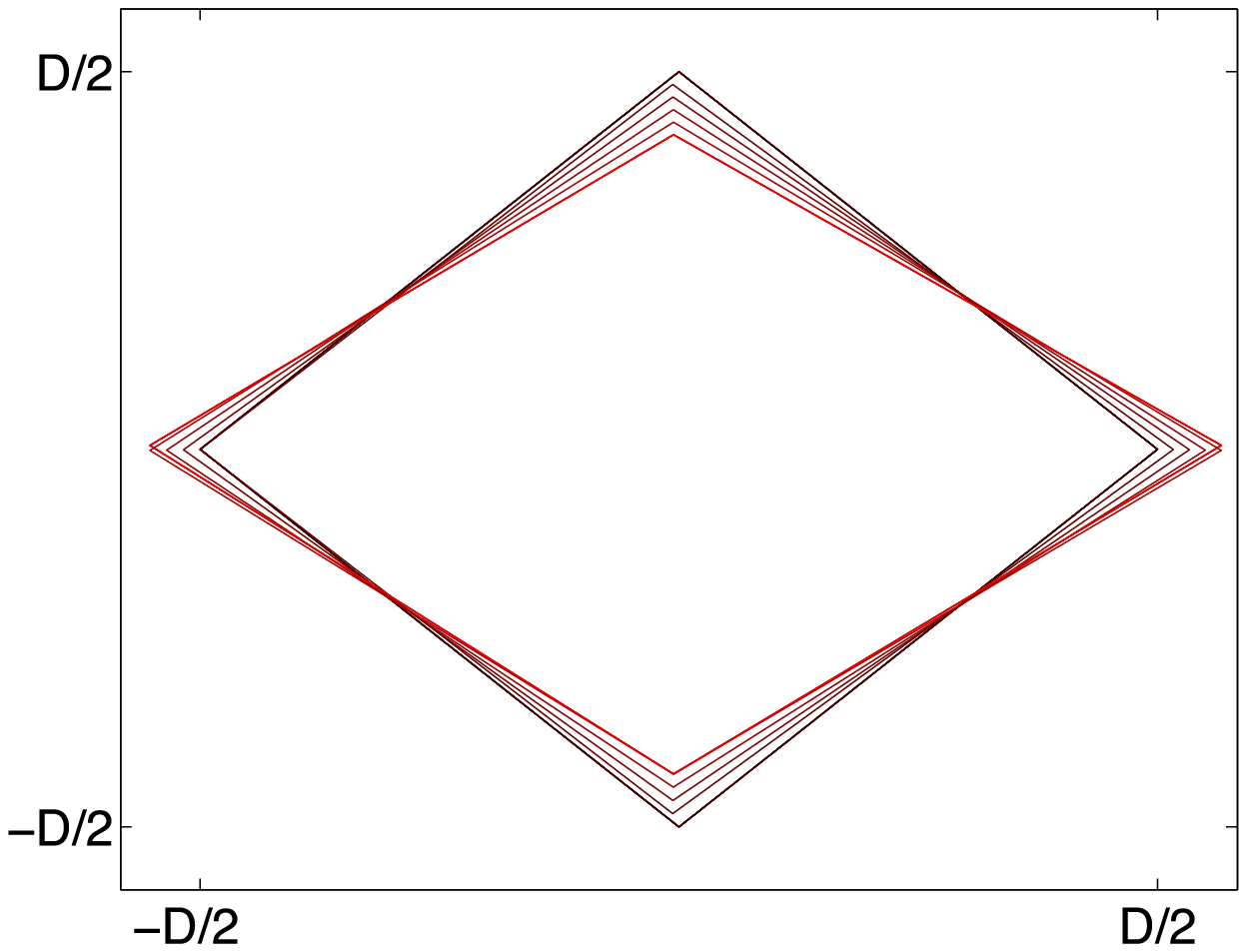}}
\centerline{\includegraphics[width=5cm,height=5cm]{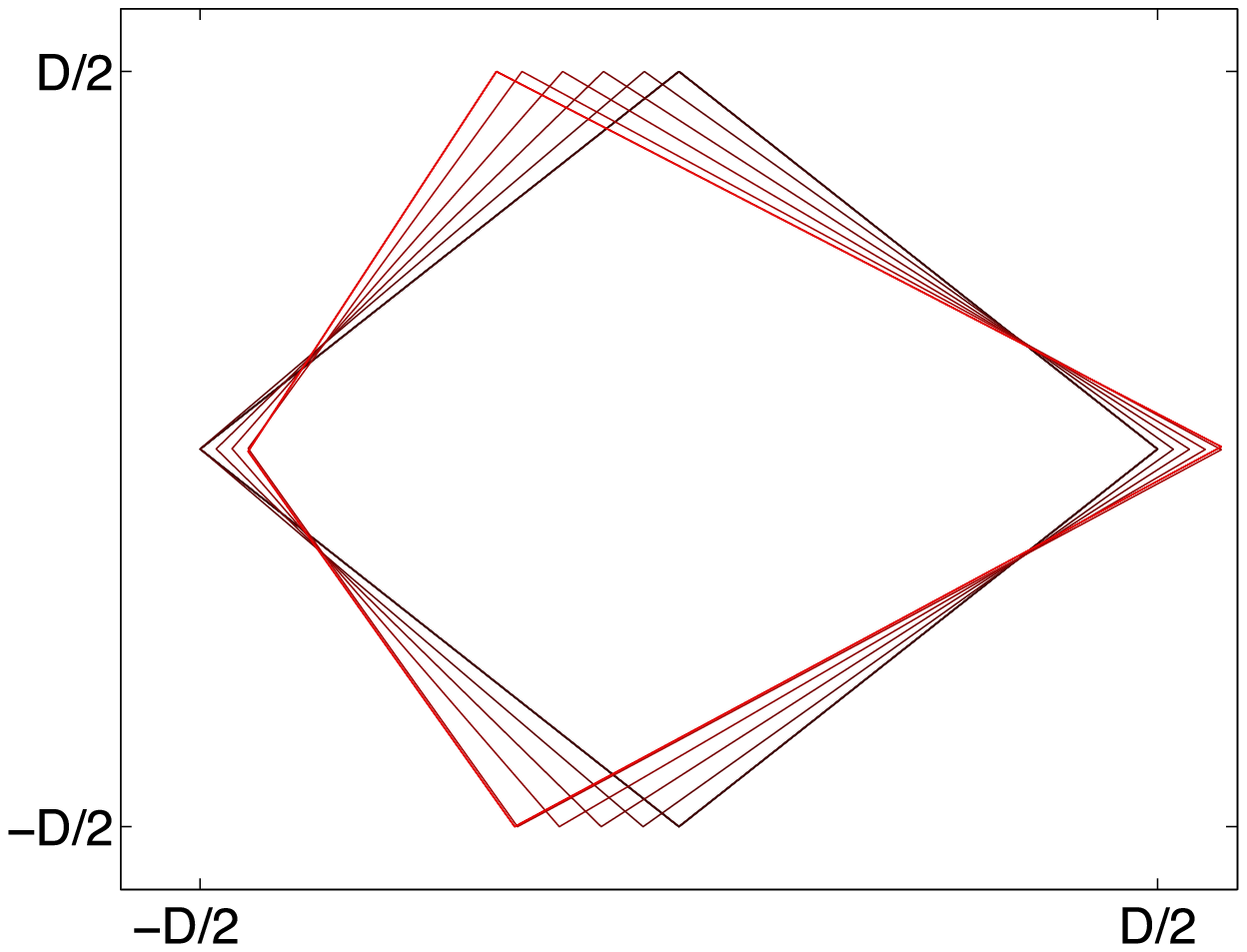},\includegraphics[width=5cm,height=5cm]{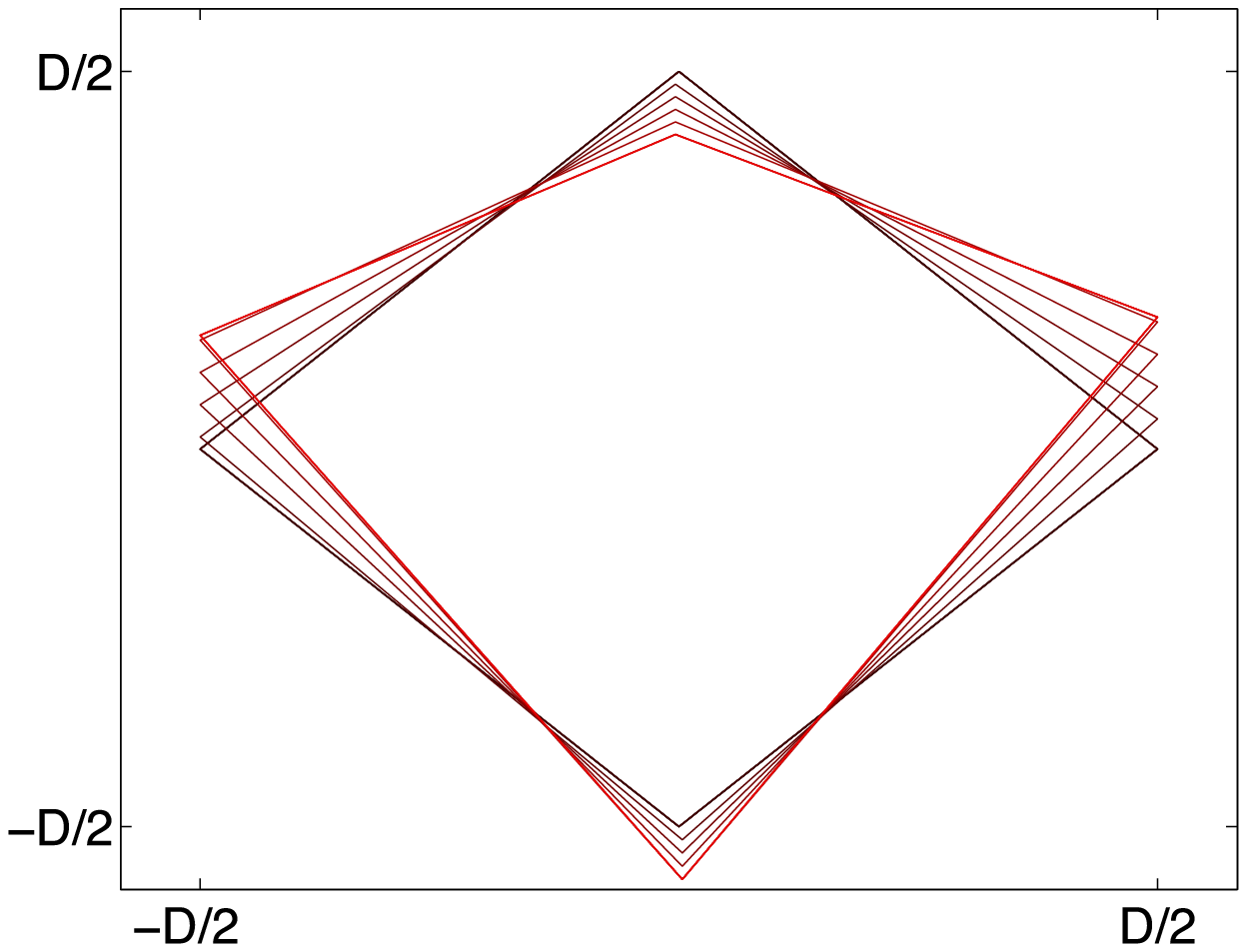}}
\centerline{\includegraphics[width=5cm,height=5cm]{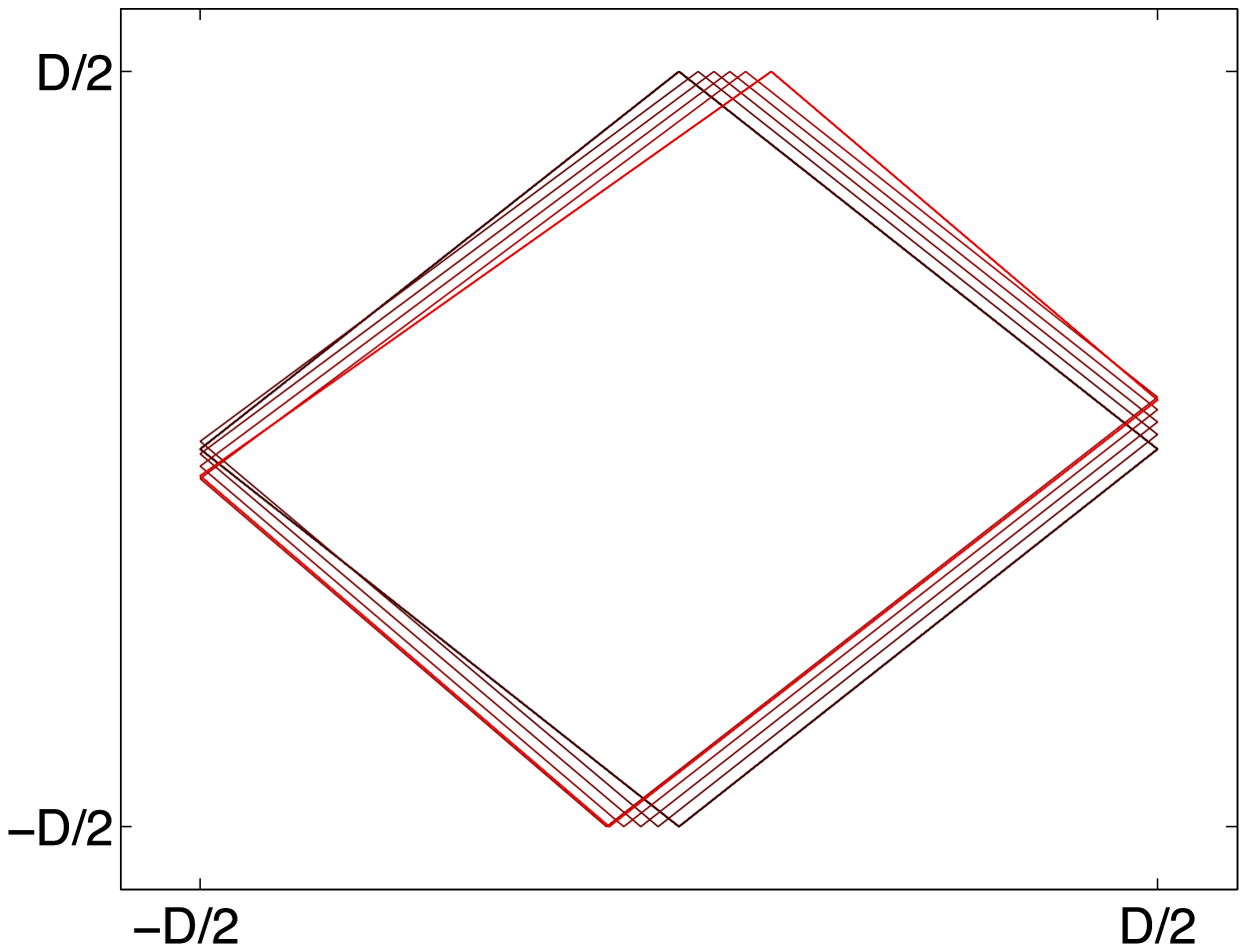},\includegraphics[width=5cm,height=5cm]{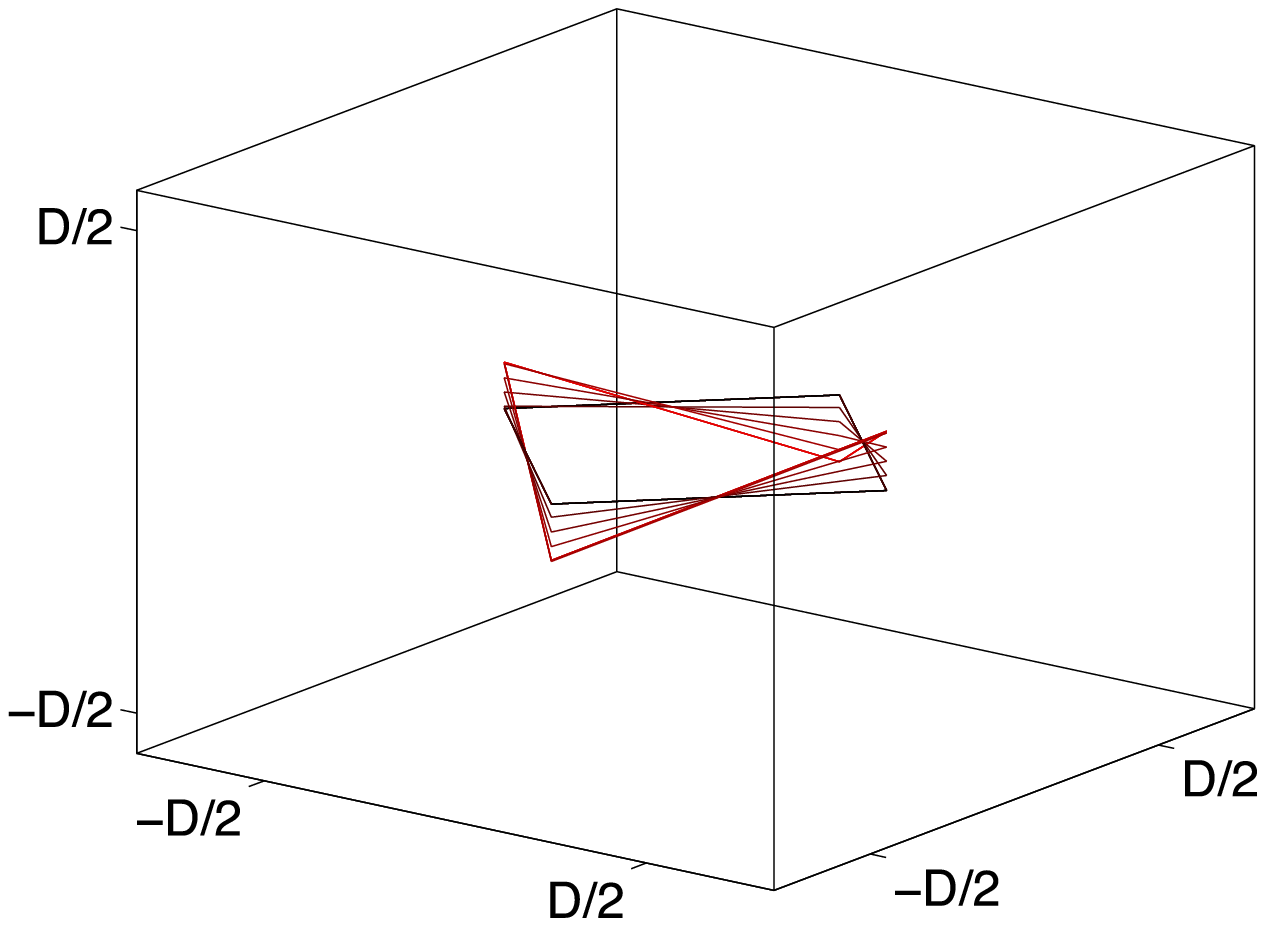}}
\caption{\label{fig:d1} Graphical representation of the optical paths related
to the non-rigid body cavity deformations calculated by imposing the Fermat's 
principle.  The black line represents the optical path of the square
cavity with diagonal length D. The red line represents the perturbed paths; the 6 deformed squares are obtained with $10^{-2}<\tau_{\alpha}/L<10^{-1}$. 
First row: diagonal common mode stretching $E_{1}$ (left); differential
mode stretching $E_{2}$ (right). Second row: shear planar deformations $E_{3}$ (left), $E_{4}$ (right). Third row: diagonal tilt $E_{5}$ (left); out of plane tilt $E_{6}$ (right).  }
\end{figure}

The geometry of the  cavity is uniquely determined by the mirror
configuration matrix $C$, representing the position of the centers
of curvature of the four mirrors. Consider a generic curve $C(t)$
in $\mathbb{R}^{3\times4}$, starting at $C_{0}$. To classify the cavity deformation we differentiate $C(t)$
with respect to $t$, and then express the components of the columns
of $\dot{C}(0)$ with respect to a suitable basis of the tangent space
in $C_{0}$
\begin{equation}
\dot{C}(0)=\sum_{\alpha=1}^{12}\tau_{\alpha}E_{\alpha}\ \nonumber ,
\end{equation}
where $E_{\alpha}$ is the $\alpha_{th}$ element of the basis of
the $3\times4$ matrices, and $\tau_{\alpha}\in\mathbb{R}$. With reference to the polygon defined by the position 
of the centers of curvature of the four mirrors, the basis
elements can be classified depending on their effects on the geometry
of the optical cavity (see figure~\ref{fig:d1}) as follows: 
\begin{itemize}
\item common and differential stretching of the diagonals
\[
 E_{1}=\frac{1}{2}\left(\begin{array}{cccc}
-1 & 0 & 1 & 0\\
0 & -1 & 0 & 1\\
0 & 0 & 0 & 0
\end{array}\right)\ ,
E_{2}=\frac{1}{2}\left(\begin{array}{cccc}
1 & 0 & -1 & 0\\
0 & -1 & 0 & 1\\
0 & 0 & 0 & 0
\end{array}\right)\ ;
\]
\item shear planar deformations
\[
E_{3}=\frac{1}{2}\left(\begin{array}{cccc}
1 & -1 & 1 & -1\\
0 & 0 & 0 & 0\\
0 & 0 & 0 & 0
\end{array}\right)\ ,
E_{4}=\frac{1}{2}\left(\begin{array}{cccc}
0 & 0 & 0 & 0\\
1 & -1 & 1 & -1\\
0 & 0 & 0 & 0
\end{array}\right)\ ;
\]
\item diagonal tilts and out of plane motions
\[
E_{5}=\frac{1}{2}\left(\begin{array}{cccc}
0 & 1 & 0 & -1\\
1 & 0 & -1 & 0\\
0 & 0 & 0 & 0
\end{array}\right)\ ,
E_{6}=\frac{1}{2}\left(\begin{array}{cccc}
0 & 0 & 0 & 0\\
0 & 0 & 0 & 0\\
1 & -1 & 1 & -1
\end{array}\right)\ ;
\]
\item rigid body translations 
\[
E_{7}=\frac{1}{2}\left(\begin{array}{cccc}
1 & 1 & 1 & 1\\
0 & 0 & 0 & 0\\
0 & 0 & 0 & 0
\end{array}\right)\ ,
E_{8}=\frac{1}{2}\left(\begin{array}{cccc}
0 & 0 & 0 & 0\\
1 & 1 & 1 & 1\\
0 & 0 & 0 & 0
\end{array}\right)\ ,
\]
\smallskip
\[
E_{9}=\frac{1}{2}\left(\begin{array}{cccc}
0 & 0 & 0 & 0\\
0 & 0 & 0 & 0\\
1 & 1 & 1 & 1
\end{array}\right)\ ;
\]
\item infinitesimal rigid body rotations 
\[
E_{10}=\frac{1}{\sqrt{2}}\left(\begin{array}{cccc}
0 & 0 & 0 & 0\\
0 & 0 & 0 & 0\\
0 & 1 & 0 & -1
\end{array}\right)\ ,
E_{11}=\frac{1}{\sqrt{2}}\left(\begin{array}{cccc}
0 & 0 & 0 & 0\\
0 & 0 & 0 & 0\\
1 & 0 & -1 & 0
\end{array}\right)\ ,
\]
\smallskip
\[
E_{12}=\frac{1}{2}\left(\begin{array}{cccc}
0 & -1 & 0 & 1\\
1 & 0 & -1 & 0\\
0 & 0 & 0 & 0
\end{array}\right)\
\]
\end{itemize}

By definition, the six distances $l_{kj}(X;C,R)$ and their functions,
such as $p$ and $\mathbf{A}$, do not depend, to the first order
in $\tau_{\alpha}$, on the rigid body translation and rotations represented
by the $E_{\alpha}$ with $\alpha=7,\dots,12$. Therefore the mirror
configuration matrix for a general deformation of the cavity is given by
\begin{equation}
C(\tau_{1},\dots,\tau_{6})=C_{0}+\sum_{\alpha=1}^{6}\tau_{\alpha}E_{\alpha}\ \nonumber .
\end{equation}
The expansion of $p^{*}(C,R)$ in power series of the $\tau_{\alpha}$,
retaining only second order terms, reads
\begin{equation}
p^{*}(C,R)=4L\left [1-\frac{\tau_{1}}{\sqrt{2}L}+\frac{\tau_{2}^{2}}{4L^{2}}
-\frac{\tau_{3}^{2}+\tau_{4}^{2}}{2L(\sqrt{2}r-2L)}  \newline
-\frac{\tau_{6}^{2}}{2L(2\sqrt{2}r-L)} \right]+o(\tau_{\alpha}^{2}). \label{eq:perNic}
\end{equation}
$p^{*}(C,R)$ depends linearly on the diagonal common stretching,
while it depends quadratically on differential stretching, shears
and tilts. Assuming that the cavity satisfies the stability condition for the confinement of
the optical rays $0\leq L/r\leq\sqrt{2}$ \cite{SalehBEA}, the coefficients
of $\tau_{3}^{2}$, $\tau_{4}^{2}$ and $\tau^{2}_6$ are negative.

Combining equation~\ref{eq:perNic} and equation~\ref{eq:arearing}, and assuming the cavity 
oriented along the direction of maximum rotational signal ($\mathrm{\mathbf{A}\parallel\mathbf{u}_{\Omega}}$),
the ring laser scale factor reads 
\begin{eqnarray}
k_{S}^{*}(C,R) & \equiv &  \frac{4}{\lambda p^{*}(C,R)}\mathbf{A}^{*}(C,R)\cdot\mathbf{u}_{\Omega} 
\nonumber  \\
& = & \frac{L}{\lambda}\Biggl[1-\frac{\tau_{1}}{\sqrt{2}L}+\frac{\tau_{1}^{2}}{2L^{2}}
-\frac{\tau_{2}^{2}}{4L^{2}}-\frac{2L+\sqrt{2}r}{4L(r-\sqrt{2}L)^{2}}(\tau_{3}^{2}+\tau_{4}^{2}) \nonumber \\
& - & \frac{L+2\sqrt{2}r}{L(4r-\sqrt{2}L)^{2}}\tau_{6}^{2}\Biggl]+o(\tau_{\alpha}^{2})\ \label{k15}.
\end{eqnarray}
From equation \ref{eq:perNic} and equation \ref{k15} we note that, to obtain an accuracy of 1 part in $ 10^{10} $ on $k_{S}$, 
the relative amplitude of $E_1$ deformation has to be $10^{-10}$, while the requirements on the other cavity deformation amplitudes 
are drastically reduced by the quadratic dependence.

\section{Discussion} 
\begin{figure}[h!!!]
\centerline{\includegraphics[width=15cm]{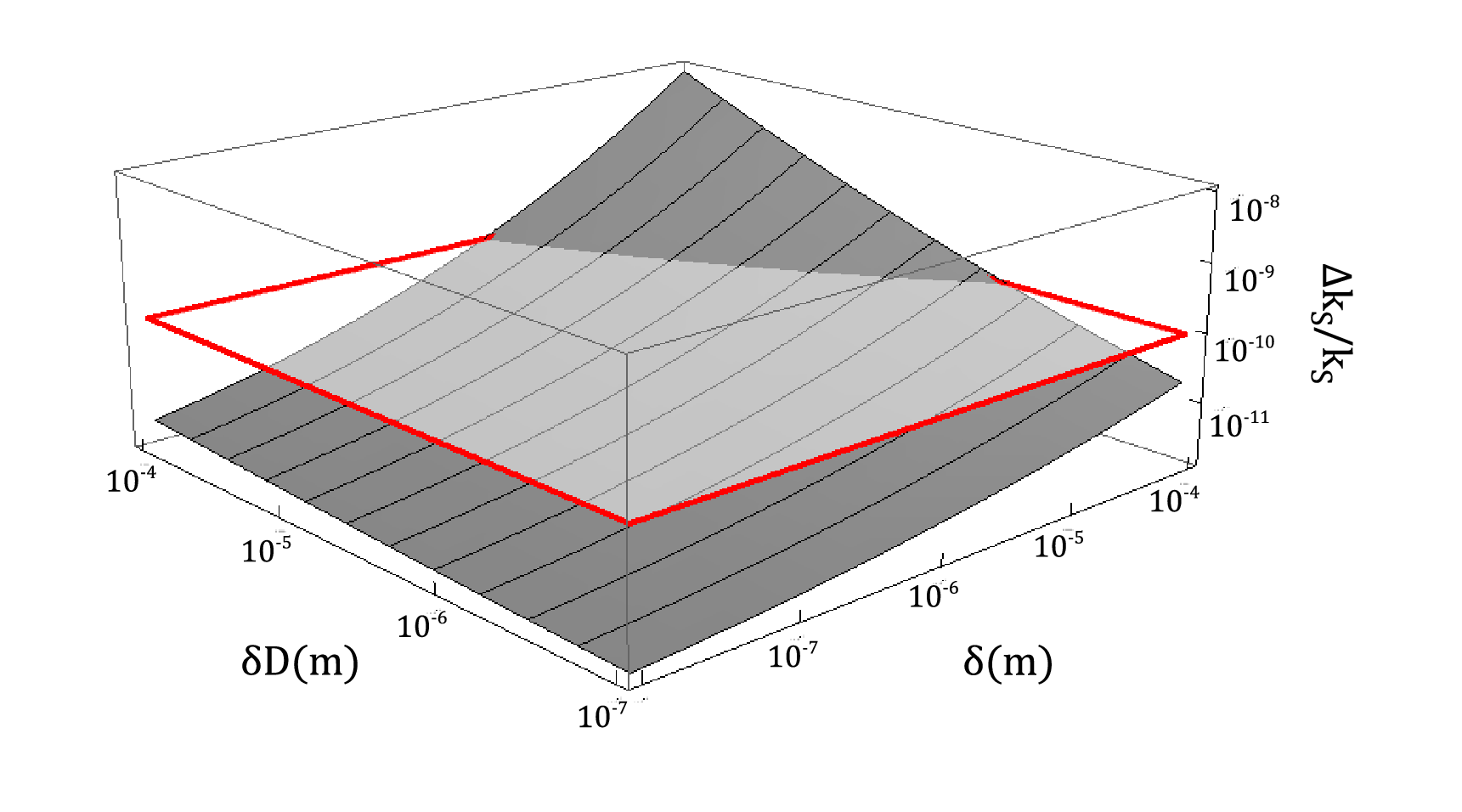}}
\caption{\label{fig:3}Relative variation of the ring laser scale factor in function of the standard deviation $\delta$ of the random displacements applied independently to each mirror position and the unbalance $\delta D$ between the two diagonal lengths. The plane $\Delta k_{S}/k{_{S}}=10^{-10}  $ (plotted with a red contour) delimits the region where the scale factor is compliant with the stability required for GINGER.}
\end{figure}

The measurement of the length of the two diagonals provides a very precise observable to constrain the $E_1$ deformation \cite{CQG}. The model has shown that the stabilization of the length of the two Fabry-P\'{e}rot resonators, and consequently of the distance between the centers of curvature of the diagonally opposite mirrors, provides a high rejection of the mirror perturbations even if their values are biased by some systematic errors. In figure~\ref{fig:3}
we report the results of a  simulation where the position of each corner mirror is varied randomly in the three directions of space 
with a standard deviation $\delta$ under the constraint of equal diagonals. 
The relative variation of the scale factor is plotted for different values of the unbalance $\delta D$ between the two diagonal absolute lengths.
Note that even in the case of a systematic error in the difference between the two diagonal cavities lengths of $10^{-6}\, \rm{m}$ \cite{CQG}, the scale factor stability is compliant with the requirements
of GINGER ($\Delta k_{S}/k{_{S}}<10^{-10}$).

 The results of equation \ref{eq:perNic} and equation \ref{k15} provide a procedure in order to approach the regular square geometry, starting from a generic mirror configuration. Let us consider  the distances $||\mathbf{c}_3-\mathbf{c}_1||$ and $||\mathbf{c}_4-\mathbf{c}_2||$ between the centers of curvature of the two pairs of opposite mirrors; imposing that the two diagonals have equal distance $d_0$, i.e.  
\begin{eqnarray}
||\mathbf{c}_3-\mathbf{c}_1||=||\mathbf{c}_4-\mathbf{c}_2||=\sqrt{(2r-\sqrt{2}L+\tau_1+\tau_2)^2+\tau_5^2}=d_0 \nonumber\\
\end{eqnarray}
we have  that $\tau_2=0$ and $\tau_1={\tau_5^2}/d_0$. \\
The radius of curvature plays also an important role: if we define $\widetilde \tau_{\alpha}=\tau_{\alpha}/d_0$, the expressions for the constrained perimeter and 
scale factor, normalized with respect to their nominal values of the square $ 4L $ and $L/\lambda $, can be rewritten as:

\begin{eqnarray}
 \widehat p^{*}(C,R) &=& 1-\frac{h}{(\sqrt{2}-2h)} (\widetilde \tau_{3}^{2} + 
\widetilde \tau_{4}^2) -\widetilde \tau_{5}^2 - \frac{\sqrt{2}h}{(4- \sqrt{2}h)} \widetilde \tau_{6}^{2}\\\nonumber
\label{eq:per2ord}
\end{eqnarray}
and
\begin{eqnarray}
\widehat k^{*}(C,R) &=& 1-\frac{h(2h+\sqrt{2})}{2(1- \sqrt{2}h)^2} (\widetilde \tau_{3}^{2} + 
\widetilde \tau_{4}^2) - \widetilde \tau_{5}^2 - \frac{2h(h+2\sqrt{2})}{(4-2 \sqrt{2}h)^2} \widetilde \tau_{6}^{2}\\\nonumber
\label{eq:k2ord},
\end{eqnarray}
where $ h=L/r$.
Now $\widehat{p^{*}}(C,R)$ and $\widehat{k^{*}}(C,R)$ are quadratic 
forms of $\widetilde \tau_{\alpha}$ with no cross terms and the regular square geometry 
corresponds to a saddle-point of $\widehat p^{*}(C,R)$. It has been already pointed out the importance
to operate the ring with a geometry as close as possible to a square, which for a large apparatus cannot be obtained by construction. 
The presence of  a saddle-point means that an experimental procedure to approach the square configuration, by monitoring the perimeter length, exists; this can be made with very high accuracy through the measurement of the laser emission frequency. The present analysis shows as well that the spot of the beams on the mirrors should be in the ideal position with tolerances of microns.\\
It is straightforward to see that the coefficients of the $\tau_\alpha$ have a very similar behaviour in both equations; in particular they are constant and equal to $1$, for $\alpha=5$; while the others $\alpha=4,6,$ depend on $h$. The ideal condition should be to work with a $h$ value so that the coefficients of $\tau_\alpha$ are large in equation 17 (which means that we can approach more accurately the regular geometry) while the corresponding coefficients in equation 18 are small (which means that the rotational sensitivity is less affected by the geometry nonregularities). By these considerations, large radius of 
curvature seams offer the best compromise. Values of $h$ around $1/\sqrt{2}$ must be avoided; actually, for $h =  1/\sqrt{2}$ the cavity becomes instable, being confocal in the sagittal plane \cite{Bilger}.

The design of the RL optical cavity should be made considering as well the features of the
resulting Gaussian beam, taking also into account the active medium inside the cavity. 
In particular, the beam radius in the waist is very interested in,
since the gain medium is usually located in it, and the optimal waist radius
should be selected to be a tradeoff between low losses (small beam radius) and high output
power (large beam radius). In this respect the model is not complete. The authors intend to make a more detailed analysis in a
subsequent work.

\section{Conclusions}
We presented a suitable formalism based on the Fermat's principle of optics 
to define the beam path geometry in a square cavity RL. This allows us to identify 
the rigid body motion of the cavity and then to classify the residual optical cavity deformations $E_{\alpha}$. 
We demonstrated that when the configuration of the mirrors is close to
that of the regular square cavity, the scale factor and the perimeter length of a RL depend linearly only on the cavity isotropic expansion 
$E_{1}$. It follows that, once the diagonal lengths are set to the same value, the effects of the
remaining deformations on the RL scale factor are quadratic in their
amplitudes $\tau_{\alpha}$. This observation is at the basis of a possible saddle-point optimization 
of the different deformations $E_{\alpha}$; keeping the mirror positions within few $\mu$m, the stability of $1$ part in $10^{10}$ in the RL scale factor $k_s$ can be obtained. The important conclusion is that an heterolithic structure can be utilized for high sensitivity and stability RL. 
\section*{Acknowledgements}
We thank A. Saccon for usefull discussions on matrix manifold theory. 
One of us (D.C.) also thanks the Eindhoven University of Technology, Netherlands, for the kind hospitality. 
Finally, the authors thank G. Cella for his constructive suggestions.\\

\end{document}